\documentstyle[aps,preprint,epsf]{revtex}
\begin{document}
\draft
\title{ Universal Parametric  Correlations of  Eigenfunctions in Chaotic and
Disordered Systems}
\author { Y. Alhassid and H. Attias}
\address {Center for Theoretical Physics, Sloane Physics Laboratory,\\
 Yale University, New Haven, CT 06520,  U.S.A.}
\date {Submitted 21 December 1994}
\maketitle

\begin{abstract}
This paper establishes the universality of parametric correlations of
eigenfunctions in chaotic and weakly disordered systems.  We demonstrate
 this universality in the framework of the gaussian random matrix process
and obtain predictions for a number of  parametric correlators, one of them
analytically.
We present numerical evidence from different models that verifies our
predictions.\\
\end{abstract}

\narrowtext

The statistical fluctuations of spectra and wave functions in complex systems
are well known to conform to the predictions of random matrix theory (RMT)
\cite{Me}. These systems, whose common feature is their non-integrability,
range from single-particle systems exhibiting chaotic behavior \cite{Bo},
such as ballistic quantum dots with an irregular confining potential
\cite{JSA}, to interacting many-particle systems, such as strongly correlated
electron models \cite{Mo} and atomic nuclei \cite{Po}. RMT predictions hold
also for electron systems with a random impurity potential which is
sufficiently weak to allow for diffusion. According to RMT, the eigenfunctions
at a given point in space are gaussian random variables \cite{Me,Pr},
whereas if the energy levels are scaled by the mean spacing, the spectral
correlations become independent of the details of the system and obey the
Wigner-Dyson spacing distribution.

It has recently been discovered that when these systems are allowed to depend
on a parameter (e.g. an external field), the correlations between spectra
belonging to different parameter values become universal upon an appropriate
scaling of the parameter \cite{SzA},\cite{SA},\cite{SZA}. The scaling factor
turned out to be the RMS of the level velocity divided by the mean spacing.

The purpose of this paper is to establish the universality of parametric
correlations of the eigenfunctions in these systems. This universality is
deduced within a general framework which we find suitable for this discussion:
the gaussian process (GP) \cite{AKM}, a random matrix process corresponding
to each one of Dyson's three gaussian ensembles (GE).   We then concentrate on
the case of
conserved time-reversal symmetry and present two parametric correlators
involving the eigenfunctions. We demonstrate the universality of these
correlators by comparing their RMT predictions, obtained from an appropriate
GP, with results of numerical simulations for a chaotic system and a disordered
system. Finally we introduce a third correlator, which we are able to calculate
analytically, and thus provide an explicit  example to the general discussion
of scaling. The universality in this case is again verified by numerical
simulations.

Dyson \cite{Dy1} showed that there are only three possible types of gaussian
ensembles that can describe a physical system, depending on its symmetry:
 orthogonal (GOE), unitary (GUE)
and symplectic (GSE). When the system depends on a
parameter $x$, and respects the same symmetry for all values of $x$, it has
been proposed \cite{AKM} that its statistical properties may be described by
corresponding gaussian processes termed respectively gaussian orthogonal
process (GOP), gaussian unitary process (GUP), and gaussian symplectic process
(GSP). A GP is a set of random $N \times N$ matrices $H(x)$ whose elements are
distributed at each $x$ according to the appropriate GE with a prescribed
correlation among elements at different values of $x$:
\begin{eqnarray} \label{1}
   \overline {H_{ij} (x) } & = & 0 \;, \nonumber \\
   \overline{ H_{ij} (x) H_{kl} (x^\prime) }  & = & {a^2 \over {2\beta}}
   f(x,x^\prime) g^{(\beta)}_{ij,kl} \;,
\end{eqnarray}
where
  $g^{(\beta =1)}_{ij,kl} = \delta_{ik} \delta_{jl} + \delta_{il} \delta_{jk}$
and $g^{(\beta =2)}_{ij,kl} = 2 \delta_{il} \delta_{jk}$.
We will be concerned with stationary processes for which the process
correlation function $f(x,x^\prime)=f(\mid x-x^\prime \mid)$ and is normalized
such that $f(0)=1$. A GP is completely determined by its first two moments
defined in (\ref{1}) and has the useful property that the joint probability
distribution of any finite number of matrices
$H(x), H(x^\prime), H(x^{\prime\prime}),\ldots$ is gaussian. In particular,
at any $x$ we have
$P\left[ H(x)\right] \propto \exp \left[ -\beta{\rm Tr}H(x)^2/2a^2\right]$. The
case $\beta=1$ corresponds to conserved time-reversal symmetry and $H(x)$ are
real symmetric, whereas for $\beta=2$ this symmetry is broken and $H(x)$ are
complex hermitean. As usual in applications of RMT we are interested in the
limit $N\rightarrow\infty$.

An example of a gaussian process is Dyson's Brownian motion model \cite{Dy2}
where $f(x)= \exp(-\gamma\mid x\mid)$. This model has been used to relate the
above-mentioned spectral correlators to the spacetime correlations of the
Sutherland-Calogero-Moser system
\cite{BNS,SLA}.

Considering now an arbitrary two-point correlation function
   $c(x,x^\prime) = \overline {{\cal O}(x){\cal O}(x^\prime)}$ where
   ${\cal O}(x)$ is some observable
involving the spectrum and eigenfunctions of $H(x)$, it can be shown
\cite{AKM,AA}   that $c(x-x^\prime)/c(0)$ depends, apart from on $a$,
 on the combination $N[1-f(x-x^\prime)]$.  To
make a correspondence with a particular system we note that $a$
sets the
mean level spacing $\Delta$ near the center of the spectrum through
$a/\Delta = \sqrt{2N}/ \pi$. Absorbing $a$ into $H(x)$ in (\ref{1}) is thus
achieved
by scaling the energies $E_i$ by $\Delta$ to get the ``unfolded'' energies
$\epsilon_i = E_i / \Delta$, leaving $c(x-x^\prime)/c(0)$ independent of $a$.
 Next,  for $x^\prime$ near $x$ we expand
\begin{eqnarray} \label{2}
   f(x-x^\prime) \approx 1 - \kappa (x-x^\prime)^2
\end{eqnarray}
and, following \cite{SA}, we consider the variance of the level velocity
$\overline {(\partial \epsilon_i / \partial x)^2}$. Writing
$\partial E_i / \partial x \approx
\langle\psi_i(x)\mid H(x^\prime)-H(x)\mid\psi_i(x) \rangle /
   (x^\prime-x)$,
where $\mid \psi_i(x) \rangle$ are the eigenstates of $H(x)$, we calculate its
variance using the joint two-matrix distribution
$P\left[ H(x), H(x^\prime)\right]$. The calculation is performed in two steps.
First, using the conditional distribution for $H(x^\prime)$ given $H(x)$ which
can be shown to be
\begin{eqnarray}\label{3}
   P\left[ H(x^\prime) \mid H(x) \right] \equiv
   P\left[ H(x), H(x^\prime)\right] / P\left[ H(x)\right] \nonumber \\
   \propto \exp \left\{ -\beta {\rm Tr}\left[ H(x^\prime) - fH(x)\right]^2
   /2a^2 (1-f^2) \right\}
\end{eqnarray}
with $f \equiv f(x-x^\prime)$, we average over $H(x^\prime)$ keeping $H(x)$
fixed to get
\begin{eqnarray} \label{4}
   \overline{\left( {\partial E_i \over {\partial x}} \right)^2} \approx
   {1 \over (x-x^\prime)^2} \left[ {a^2 \over \beta}(1-f^2)+(1-f)^2 E_i(x)^2
   \right] \;.
\end{eqnarray}
Second, taking the limit $x^\prime \rightarrow x$ and using (\ref{2}),
the second term on the r.h.s. of (\ref{4}) vanishes and we obtain the following
expression for the non-universal quantity $\kappa$:
\begin{eqnarray}\label{5}
   \kappa = \beta \frac{\pi^2}{4} \frac{1}{N} \overline{ \left(
   \frac{\partial\epsilon_i }{ \partial x}  \right)^2 } \;.
\end{eqnarray}
Thus after the scaling
\begin{eqnarray}\label{6}
   x\rightarrow\bar{x} =
   \left[ \overline{ (\partial\epsilon_i / \partial x)^2}\right]^{1/2} x
\end{eqnarray}
we get
\begin{eqnarray}\label{7}
   f\approx 1 - \beta {{\pi^2}\over 4}
   {{(\bar{x}-\bar{x}^\prime)^2} \over N} \;,
\end{eqnarray}
 and all correlators, being determined by the combination $N(1-f)$,
 become universal as  functions of $\bar{x}-\bar{x}^\prime$.

A few remarks are in order. First, the scaling (\ref{6}) is identical to that
found in \cite{SA} for spectral correlators. Second, this scaling was derived
here under the assumption that the second derivative of $f$ is the first
non-vanishing one (see (\ref{2})). This is not always the case  as exemplified
by  Dyson's
Brownian motion model for which $f \approx 1-\gamma\mid x\mid$.  The more
 general case is discussed elsewhere  \cite{AKM}.
 Third, the form (\ref{7})
of $f$ implies that the typical correlation length $x$ scales like
${1 \over {\sqrt N}}$ in the GP, as was shown in \cite{SLA} and will be seen
explicitly in the analytical expression for the correlator
    $\tilde o(\omega ,x-x^\prime)$ derived below. Fourth, and most importantly,
 we note that previous treatments of parametric correlations had to invoke the
supersymmetry method \cite{SA} or Dyson's Brownian motion model \cite{BNS}
and could demonstrate universality of spectral correlators only  (however,  see
\cite{TAA}). In the GP framework, on the other hand, the universality of all
correlators emerges quite simply. In particular, correlators involving the
eigenfunctions are universal.

We shall demonstrate the last point by investigating two such quantities,
the averaged parametric overlap
\begin{eqnarray}\label{8}
   o(x-x^\prime)  =
 \overline{\mid \langle \psi_i(x) \mid \psi_i(x^\prime)
   \rangle \mid ^2}
\end{eqnarray}
and the projection correlator
\begin{eqnarray}\label{9}
   p(x-x^\prime) & = & \overline { \langle \phi \mid \psi_i (x) \rangle
   \langle \phi \mid \psi_i (x^\prime) \rangle } \;.
\end{eqnarray}
Both measure the decorrelation of wave  functions as their separation along $x$
increases. $o(x-x^\prime)$ gives the overlap of wavefunctions at different $x$,
whereas $p(x-x^\prime)$ provides the correlation between their components along
a fixed normalized vector $\mid \phi \rangle$. To see that $p(x-x^\prime)$ is
independent of the
choice of $\mid \phi \rangle$, notice that had we chosen instead
$\mid \phi^\prime \rangle = U\mid\phi\rangle$ for some
unitary $U$, we could have rotated $H(x)$ by $U^\dagger$ at each $x$
without affecting the probability measure (see (\ref{3})), thereby recovering
the original result for $p(x-x^\prime)$. In particular, for
    $ \; \mid\phi \rangle = \mid r \rangle$ the projection correlator describes
the correlation between eigenfunctions belonging to different values of $x$ at
a given space point $r$:
   $p(x-x^\prime) = \overline {\psi^x_i(r) \psi^{x^\prime}_i(r)}$.
For the orthogonal case $\psi^x_i(r)$ are real
and determined up to a sign, which is easy to keep fixed as $x$ varies.

We cannot provide analytical expressions for these correlators. However, since
they are universal we can use any gaussian process to obtain a theoretical
prediction for them from a random matrix simulation. A simple GOP is generated
by $H(x) = H_1 \cos x  + H_2 \sin x$,  where $H_1, H_2$ are independent
GOE matrices. This process is stationary with
$f(x-x^\prime)= \cos{(x-x^\prime)}$  and the scaling (\ref{6}) is readily found
to be $x \rightarrow \bar{x}=(\sqrt{2N}/\pi ) x$. The theoretical curves,
generated by a simulation of random matrices with $N=150$, are given by the
dashed lines in Fig. \ref{fig1}.

To verify the universality we studied $o(x-x^\prime)$ and $p(x-x^\prime)$
in both  a chaotic system and a disordered system. The first is the
interacting boson model (IBM), a many-body system used to describe collective
states of medium and heavy mass nuclei \cite{IA}. Its constituents are bosons
which model nucleon pairs coupled to angular momentum of 0 or 2. Depending on
two parameters $\chi, \eta$ the IBM can be integrable or non-integrable. The
time-dependent mean field equations, obtained in the limit of an infinite
number of bosons and which constitute the classical limit, are correspondingly
regular or chaotic \cite{ANW}. We have calculated the above correlators as a
function of $\chi$ in the regime $\eta=0$, $ -0.8 < \chi  < -0.5$ where the
system is almost fully chaotic. Since the total spin $J$ of the nucleus is a
conserved quantum number,  we can study the correlations for different values
of the spin. Results for $J=2$ and $J=6$ with 25 bosons are displayed in the
right panel of Fig. \ref{fig1}.

The second system we studied was the Anderson model, a two-dimensional lattice
Hamiltonian with on-site disorder and nearest-neighbor hopping. The site
energies $W_i$, measured in units of the hopping term, are uniformly
distributed in $\left[-W/2,W/2\right]$ where $W$ controls the transition
between the diffusive and localized regimes. We considered the cases of
cylindrical geometry with $W=4$ and toroidal geometry with $W=2$, introducing a
parametric dependence by adding a step potential of strength $x$ along one of
the lattice directions \cite{SA}. Results for a $27 \times 27$ lattice are
presented on the left panel of Fig. \ref{fig1}. Both correlators in both
systems are in excellent agreement with the GOP prediction.

The last part of this paper introduces a third correlator which we are
able to calculate analytically using Efetov's supersymmetry method \cite{Ef}.
This correlator, related to the parametric overlap (\ref{8}) but involving both
energies and eigenfunctions, is given by
\begin{eqnarray}\label{10}
   \tilde{o}(\omega, x-x^\prime)  =
      \frac{\sum_{ij} \overline{\mid \langle
   \psi_i(x)\mid \psi_j(x^\prime) \rangle \mid^2 \delta\left( \epsilon_i(x)+
   \omega - \epsilon_j(x^\prime)\right)}}
   {\sum_{ij} \overline{ \delta\left( \epsilon_i(x)+\omega -
   \epsilon_j(x^\prime) \right) }} \;.
\end{eqnarray}
It measures the averaged parametric overlap of eigenfunctions whose
corresponding energies are separated by $\omega$ in units of the mean spacing.
We now outline the calculation of the numerator of (\ref{10}), denoted
$o(\omega,x-x^\prime)$ (details will be published elsewhere \cite{AA}). In
order to employ the supersymmetry method we first express it in terms of Green
functions:
$o(\omega,x-x^\prime) = \frac{1}{2\pi^2}  {\rm Re} {\rm Tr} \left[
 \overline{G(\epsilon^-,x) G(\epsilon^+ +\omega,x^\prime)} -
 \overline{G(\epsilon^-,x) G(\epsilon^- +\omega,x^\prime)} \right]$
where $\epsilon^\pm = \epsilon\pm i\delta$.
Each Green function is then written as an integral over $4N$-dimensional graded
vector $\Psi$. After performing the GOP averaging, a Hubbard-Stratonovich
transformation and a subsequent integration over the $\Psi$-variables result
in an integral over a $16$-dimensional graded matrix $R$. This integration is
carried out in the saddle-point approximation, which is exact in the
infinite-$N$ limit. In this limit the quadratic corrections decouple and we are
left with an integral over the saddle-point manifold:
\begin{eqnarray}\label{11}
   o(\omega,x-x^\prime) = {1 \over {2\Delta^2}}{\rm Re}
   \int D\left[Q\right] P\left[Q\right] \exp\left( F\left[Q\right]\right)
\end{eqnarray}
where $Q$ is a $8 \times 8$ graded matrix and
\begin{eqnarray}\label{12}
   P\left[Q\right] & = & {\rm Trg}(QS^{12}_{bb}) {\rm Trg}(QS^{21}_{bb}) \;,
   \nonumber \\
   F\left[Q\right]& = & i{\pi\over 4} (\omega +2i\delta) {\rm Trg}(Q\Lambda )
      +{1\over 16}N\kappa (x-x^\prime)^2 {\rm Trg}\left[ Q,\Lambda\right]^2
   \;.
\end{eqnarray}
Using Efetov's parametrization of $Q$ \cite{Ef} most of the integrals are
elementary and the result is given by
\begin{eqnarray}\label{13}
   o(\omega,x-x^\prime) = {1 \over {2\Delta^2}}{\rm Re}
   \int D\left[\lambda\right] P\left[\lambda\right] \exp(F\left[\lambda\right])
   \;,
\end{eqnarray}
where
\begin{eqnarray}\label{14}
   \int D\left[\lambda\right] & \equiv & \int_1^\infty \int_1^\infty
\int_{-1}^1
   {{(1-\lambda^2) d\lambda_1 d\lambda_2 d\lambda} \over
    {(\lambda_1^2+\lambda_2^2+\lambda^2-2\lambda_1\lambda_2\lambda -1)^2}} \;,
   \nonumber \\
   P\left[\lambda\right]  &= & 2\lambda_1^2\lambda_2^2-\lambda_1^2
   -\lambda_2^2-\lambda^2+1 \; , \nonumber \\
   F\left[\lambda\right]  & = & i\pi (\omega +2i\delta)(\lambda_1\lambda_2
   - \lambda)
   -N\kappa (x-x^\prime)^2
   (2\lambda_1^2\lambda_2^2-\lambda_1^2-\lambda_2^2-\lambda^2+1) \;.
\end{eqnarray}
Note the combination $N(x-x^\prime)^2$ in $o(\omega,x-x^\prime)$ which
explicitly suggests the $\sqrt{N}$ scaling of $x$ mentioned above. The scaling
(\ref{6}) removes the dependence on the system-specific $\kappa$ through the
substitution
 $N\kappa (x-x^\prime)^2 = {{\pi^2}\over 4} (\bar{x} - \bar{x}^\prime)^2$.
Our result for the denominator of (\ref{10}), after scaling, becomes identical
to the level density correlator calculated in \cite{SA}. An expression for
$\tilde o(\omega,x-x^\prime)$ can similarly be derived for the case of broken
time-reversal symmetry by GUP-averaging the Green function product \cite{AA}.

We verified our derivation of $\tilde o(\omega,x-x^\prime)$  by a comparison
with a GOP simulation and confirmed the universality of this correlator by
studying it in the above two cases of the Anderson model. We remark that the
expression in (\ref{13})-(\ref{14}) corresponds to a regularization of
(\ref{10}) by convoluting both the numerator and the denominator with a
Lorentzian of width $\delta$. The same operation, which amounts to smearing the
$\delta$-functions, was performed in the numerical computations. The results
are displayed in Fig. \ref{fig2} for $\delta = 0.05$ in units of the mean
spacing and various values of $\omega$ and are in excellent agreement with the
GOP prediction.

  In conclusion,  we  have shown that the concept of the gaussian
process is particularly suitable for a discussion of parametric correlations
in chaotic and disordered systems and for demonstrating their universality.
We established the existence of universal parametric correlations of
eigenfunctions in these systems, provided predictions for three such
correlators, and verified them in different models.

This work was supported in part by the Department of Energy Grant
DE-FG02-91ER40608.

\begin{figure}
\caption { The eigenfunction correlators $o(\bar{x}-\bar{x}^\prime)$ (top) and
$p(\bar{x}-\bar{x}^\prime)$ (bottom) as a function of the scaled parameter
(see Eqs. (6),(8) and(9)).
The dashed lines are the GOP prediction obtained from simulations of
$H(x) = H_1\cos x + H_2\sin x$ with $N=150$, using the middle third of the
spectrum. On the left are calculations in the IBM in its chaotic regime,
including $80$ states out of $117$ for $J=2$ (pluses) and $130$ states out of
$184$ for $J=6$ (squares). On the right are calculations in the Anderson model
in its diffusive regime, including the middle $200$ states out of $729$.
Results are given for a cylindrical geometry with $W=4$ (circles) and for a
toroidal geometry with $W=2$ (crosses).}
\label{fig1}

\vspace{5 mm}

\caption{The eigenfunction correlator
$\tilde o(\omega,\bar{x}-\bar{x}^\prime)$ (see Eq. (10)) as a function of the
scaled parameter for several values of $\omega$ measured in units of the mean
spacing. The solid lines are the analytical results (see Eqs. (13)-(14)), the
dashed lines are the GOP simulations, and the circles and crosses are the
results for the same cases of the Anderson model as in Fig. 1. The
$\delta$-functions in (10) are regularized with a Lorentzian of width
$\delta=0.05$ in units of the mean spacing. }
\label{fig2}
\end{figure}

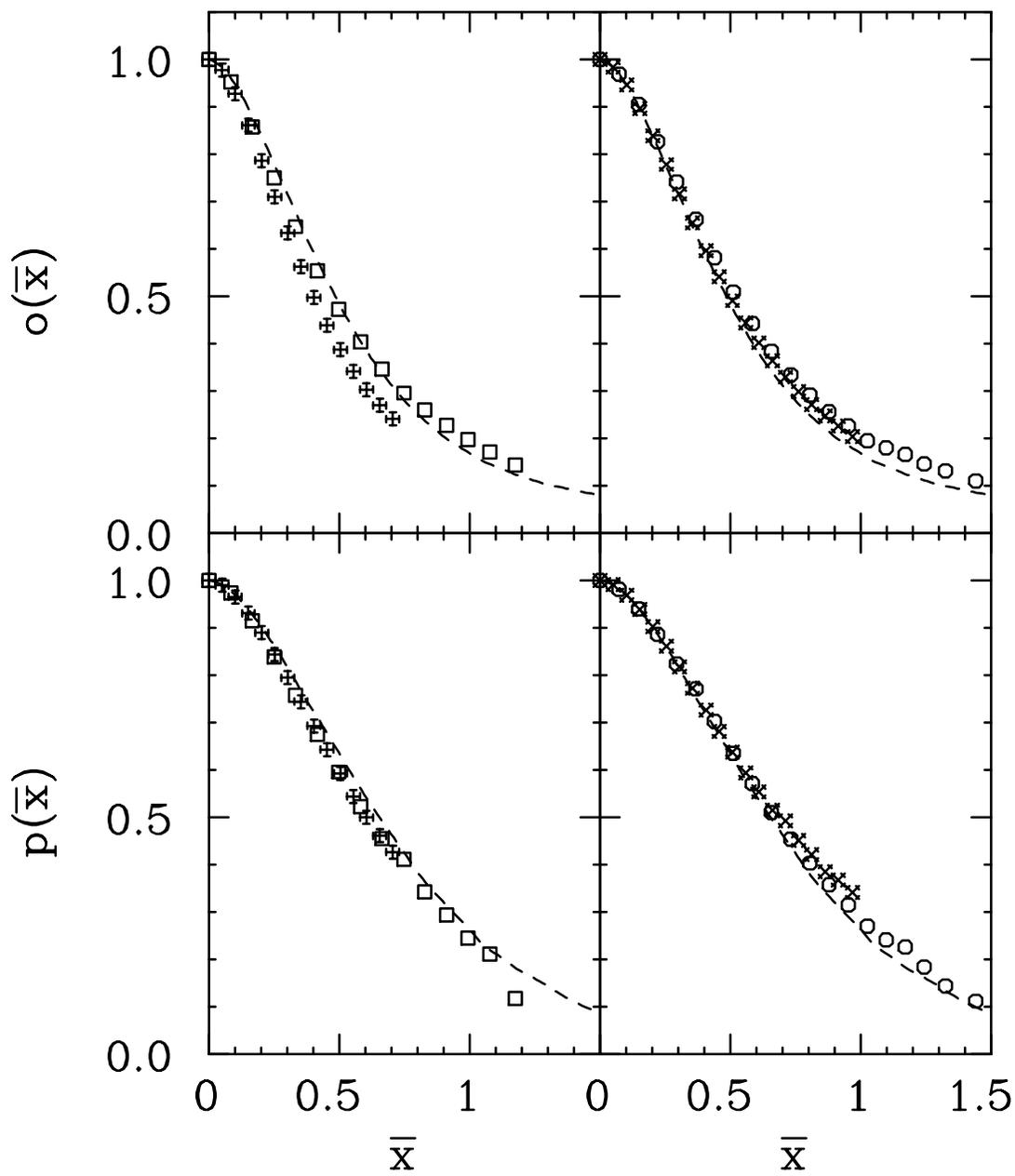
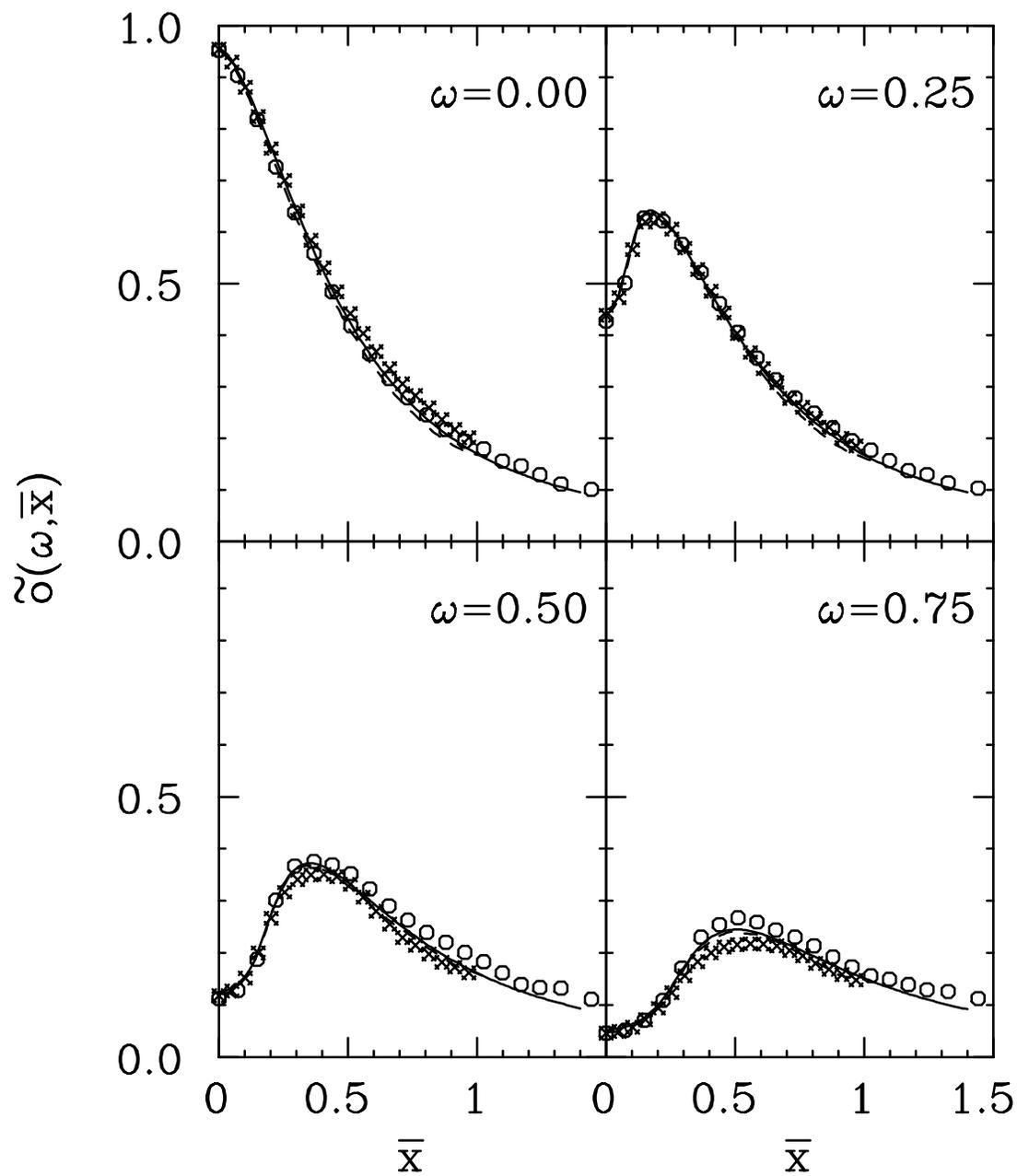
\end{document}